\documentclass[10pt, paper=a4, UKenglish]{article}
\usepackage{graphicx}
\usepackage{balance}
\usepackage[font={small,it}]{caption}
\usepackage[utf8]{inputenc}
\usepackage[T1]{fontenc}
\usepackage{graphicx}
\usepackage{epsfig}
\usepackage[english]{babel}
\usepackage{listings}
\usepackage{hyperref}
\usepackage{authblk}
\usepackage{fancyhdr}
\usepackage{floatflt}
\usepackage{float}
\usepackage{caption}
\usepackage{wrapfig}
\usepackage{rotating}
\usepackage{longtable}
\usepackage{chngcntr}
\usepackage{relsize}
\usepackage{amsmath}
\usepackage{amsfonts}
\usepackage{amssymb}
\usepackage[amssymb, cdot, mediumqspace, thickqspace]{SIunits}
\usepackage{mathrsfs}
\usepackage{booktabs}
\usepackage{braket}
\usepackage{appendix}
\usepackage{pdfpages}
\usepackage{placeins}
\usepackage{titlesec}
\usepackage{mathtools}
\usepackage{enumitem}
\usepackage{color}
\usepackage{mhchem}
\usepackage{slashed}
\usepackage{feynmp}
\usepackage{feynmp-auto}
\usepackage{lineno}
\usepackage{ulem}
\usepackage[bottom]{footmisc}
\usepackage{tabularx}
\usepackage{graphicx,lipsum}
\usepackage{tikz} 
\usetikzlibrary{shapes,arrows}
\usepackage[numbers,sort&compress]{natbib}
\definecolor{lightblue}{rgb}{0.145,0.6666,1}

\def\Title#1{\begin{center} {\Large #1 } \end{center}}
\def\Author#1{\begin{center}{ \sc #1} \end{center}}
\def\Address#1{\begin{center}{ \it #1} \end{center}}

\newcommand\pubblock{\rightline{\begin{tabular}{l} Proceedings of the CTD/WIT 2019\\ \pubnumber\\
         \pubdate  \end{tabular}}}

\newenvironment{Abstract}{\begin{quotation} \begin{center} 
             \large ABSTRACT \end{center}\bigskip 
      \begin{center}\begin{large}}{\end{large}\end{center} \end{quotation}}

\newenvironment{Presented}{\begin{quotation} \begin{center} 
             PRESENTED AT\end{center}\bigskip 
      \begin{center}\begin{large}}{\end{large}\end{center} \end{quotation}}

\def\beq{\begin{equation}}
\def\eeq#1{\label{#1}\end{equation}}
\def\eeqn{\end{equation}}

\def\beqa{\begin{eqnarray}}
\def\eeqa#1{\label{#1}\end{eqnarray}}
\def\eeqan{\end{eqnarray}}

\let\bar=\overbar

\def\Dslash{\not{\hbox{\kern-4pt $D$}}}
\def\dslash{\not{\hbox{\kern-2pt $\del$}}}

\def\msb{{\bar{\ssstyle M \kern -1pt S}}}

\usepackage{color}
\usepackage{lineno}
\usepackage{subfig}
\usepackage{hyperref}

\newcommand\pubnumber{PROC-CTD19-020 \\ CMS CR-2019/068}
\newcommand\pubdate{\today}

\def\affiliation{
On behalf of CMS Collaboration, \\
Scuola Normale Superiore and INFN Sez. di Pisa, Italy}

\newcommand{\conference}{Connecting the Dots and Workshop on Intelligent Trackers (CTD/WIT 2019)\\
Instituto de F\'isica Corpuscular (IFIC), Valencia, Spain\\ 
April 2-5, 2019}

\usepackage{fancyhdr}
\definecolor{mygrey}{RGB}{105,105,105}
\graphicspath{{figure/}}

\begin{document}


\large
\begin{titlepage}
\pubblock

\vfill
\Title{DeepCore: Convolutional Neural Network for high $p_T$ jet tracking}
\vfill

\Author{Valerio Bertacchi}
\Address{\affiliation}
\vfill

\begin{Abstract}
Tracking in high-density environments, such as the core of TeV jets, is particularly challenging both because combinatorics quickly diverge and because tracks may not leave anymore individual “hits” but rather large clusters of merged signals in the innermost tracking detectors. In the CMS collaboration, this problem has been addressed in the past with cluster splitting algorithms, working layer by layer, followed by a pattern recognition step where a high number of candidate tracks are tested. Modern Deep Learning techniques can be used to better handle the problem by correlating information on multiple layers and directly providing proto-tracks without the need of an explicit cluster splitting algorithm. Preliminary results will be presented with ideas on how to further improve the algorithms.
\end{Abstract}

\vfill

\begin{Presented}
\conference
\end{Presented}
\vfill
\end{titlepage}
\def\thefootnote{\fnsymbol{footnote}}
\setcounter{footnote}{0}
%

\normalsize 


\section{Introduction and motivations}

The events with high-energy (trasverse momentum $p_T^\text{jet}\gtrsim 0.5$ TeV) jets emission are part of a rich physics program at LHC, both for the New Physics searches and the Standard Model (SM) physics. The boosted environment improves the performances of the analysis which involves high-mass SM objects, like vector bosons or b-quarks \cite{CMS:bjet,CMS:substructures,LHC:substructures}. The track reconstruction inside the jet is a fundamental step for all the analyses which want to investigate the composition of the jets, looking for substructures and specific particle signatures. In the CMS experiment the full reconstruction of the event relies on the Particle Flow algorithm \cite{CMS:ParticleFlow}, which smartly combine the information of the subdetectors to assign to each reconstructed object a particle tag, rebuilding the entire event. The silicon tracker information is one of the blocks of the Particle Flow, and improvement in tracking gives large benefits to the entire event reconstruction of CMS.

The CMS experiment is an electromagnetic spectrometer composed of a superconducting solenoid which provides a magnetic field of 3.8 T. Within the solenoid volume there is, from the interaction point (IP) to outside, the silicon pixel and strip tracker, the lead tungstanate electromagnetic calorimeter, the hadronic calorimeter composed by alternated layer of brass and scintillators. The muon detector is composed of gas chambers embedded in the steel yoke outside the superconducting solenoid. A more detailed description of the CMS detector, together with a definition of the coordinate system used and the relevant kinematic variables, can be found in Ref.~\cite{CMS:CMS}. In particular, the pixel detector  \cite{CMS:pixel} is composed of four layers in the barrel region ($\eta\lesssim 1.4$) and three disks in the endcap region (which offer a 4-hits coverage up to $\eta= 2.5$). The radii of the barrel layer are 29, 68, 109, 160~mm, the distances of the disks from the IP are 291, 396, 516 mm. The pixels size is $100\times 150$~$\mu$m$^2$ in both regions. The resolution is order of 10~$\mu$m in $r-\phi$ and 25~$\mu$m in $z$ directions. 

The CMS track reconstruction algorithm, called \textit{Combinatorial Track Finder} (CTF), is based on the combinatorial Kalman filter (CKF)\cite{fruhwirth:tracking,billoir:ckf,billoir:ckf2,mankel:ckf}. The two main ideas of this approach are: to perform pattern recognition and track fitting in the same framework and to manage the high level of complexity of the events (i.e. the combinatorial burden) with multiple passes of the same reconstruction sequence. The first iterations look for tracks which are easier to find and then they remove the associated hits. Then the following iterations look for more difficult kinematic regions (low or very high $p_T$, displaced vertex, high $\eta$ \dots), but each iteration search in less dense environment because of the removed hits. This process is called \textit{iterative tracking} and each pass proceeds in four steps:

\begin{enumerate}
\item \textbf{Seed generation}. Building of proto-tracks with the use of few hits (form 2 to 4) from specific layers of the tracker. This rough estimation of track parameters will be used as starting point for the second step. The minimum requirements to obtain an estimate of the tracks parameters are three points, as two 3D hits with the vertex constraint or three 3D hits. 
For each iteration a set of \textit{seeding layers} and a \textit{tracking region} are defined: the seeding layers are the detectors where the seed hits are searched (a pair, triplet or quadruplet of tracker layers), the tracking regions are  the kinematic or geometric selection applied on the hits to identify the seeds to define the phase space of region of interest. If the seeding layers are three or four pixel detector layers a Cellular Automaton \cite{pantaleo:CA} is used to produce the seed list instead of the tracking region constraint.   
\item \textbf{Pattern Recognition}. Extrapolation from the track-seeds to the outer layers of the tracker looking for compatible hits, exploring multiple hypotheses. The extrapolation is done taking into account the material effects (multiple scattering, energy losses), first moving outward and then repeating the extrapolation inward to recover precision in the seeding region.
\item \textbf{Fitting}. Fitting using the Kalman filter and smoother \cite{fruhwirth:kalman}, moving outward with the Runge-Kutta propagator \cite{runge-kutta} which takes into account both the material effects and the inhomogeneities of the magnetic field.
\item \textbf{Quality flagging}. Flagging the track candidates with different tag depending on their quality (based on number of hits, $\chi^2$, track parameters \dots), or discarded if the quality results too low. 
\end{enumerate}
After the selection, the track collections from the various iterations are merged in a single collection, called \textit{general tracks}. More details about the track reconstruction can be found in Ref.~\cite{CMS:tracking}. 

The number of charged particle tracks and their spatial density inside the jets grows with the energy of the jet and dedicated iteration for high energy jet was added in 2015, because the tracking performance in the jet core (i.e. the central region) result lower than the average \cite{rizzi:jetcore}. This iteration, called \textit{jetCore} was added as last of the iterative tracking and searches seeds only in a cone of $\Delta R=\sqrt{\Delta\phi^2+\Delta\eta^2}<0.1$ around the jet axis (from calorimeter deposit) if $p_T^\text{jet}>100$ GeV. The seeds are built with pairs of hits on the pixel detector and/or in the internal strip detector barrel, compatible with $p_T>10$ GeV, with the vertex constraint. In addition, the CKF tests a larger number of candidates in the jet core cone region ($\sim 50$ against the standard 5). The tracks in the jet core, due to the high density, often leave on the pixel layers large merged cluster and not individual hits. A dedicated \textit{k-means} \cite{k-means} based cluster splitter was developed to face the merged clusters, which exploits the jet axis information to predict the cluster shape and charge for a single particle cluster or a multiple-particle merged cluster.
The performance of the jetCore iteration (called from now standard jetCore) are suboptimal: the jetCore iteration improves the total tracking efficiency, but a simulation with an ideal cluster splitting reveals that there is still room for improvement. 
Anyway, tracking efficiency still degrades in the jet core also with the ideal splitting, point out that the inefficiency is not due to the merged cluster only. Therefore has been decided to change approach and develop a new version of the jetCore seeding algorithm avoiding an explicit splitting step and using the combined information of multiple pixel detector layers to produce a new list of jetCore seeds, instead of focusing to improve a layer-by-layer cluster splitting. An Artificial Neural Network, called \textit{DeepCore}, has been developed, properly trained and tested in CMS reconstruction software to cope with this task. The description of DeepCore together with its performance is presented in the following sections.

\section{Description of the DeepCore network}

In this section the general strategy on which the novel high $p_T$ jet seeding algorithm is based is presented. Then the details of DeepCore are described and in the last part the integration of the network in the CMS CTF is shown. DeepCore is currently developed for barrel region only, therefore the pixel detector is simply made of four cylindrical layers in this framework.

\subsection{The strategy}

The purpose of the seeding algorithm is to produce a list of track-seed i.e. sets of track parameters for the interested tracking region. The primary goal of the algorithm is to find additional seeds in the jet core region, recovering seeding efficiency lowering the fake tracks rate. This result can be reached producing better quality seeds in term of track parameters. The secondary goal is to lower the time consumption of jetCore iteration, currently one of the most expensive of the CTF (because of the large number of explored candidates).

Because the previous cluster splitting algorithm resulted suboptimal this explicit step has been skipped. The seeding algorithm produces directly the list of seeds (i.e. track parameters) from the raw pixel detector information, without any clustering algorithm on the top. A good candidate to reproduce the function 
\[
\vspace{-0.2 cm}
f : \{\text{raw pixel information}\} \longrightarrow \{\text{list of track seeds}\} 
\]
is an Artificial Neural Network (NN). With \textit{raw pixel} information from now on it is referred to individual pixel charge and position, without any clustering algorithm, but the default charge calibration and zero-suppression algorithms  applied. 

In the wide field of NN a Convolutional Neural Network (CNN) has been used to face the problem. The CNNs \cite{CNN} are one of the most natural choices with a 2D-picture input, like the pixel detector information. Each node of the network can be interpreted as a single pixel, each node of the hidden layers is connected only to few nodes of the previous layer i.e. receives information only from a small region of the layer. The values of the previous layer inside the region are combined with a specific \textit{filter} to produce the weight of the node of the hidden layer. The CNN swipes the filter along the entire 2D input looking for common features in the layer sharing properly the weights. The CNN uses multiple filters to increase the feature-discovery power, exploring multiple times the entire layer. The relevant parameters are the number of filters (how many kinds of features are expected), the dimension of the filters (how many pixel are needed to identify a feature) and of the number of convolution layer (the complexity of the features).

In the tracking environment the pixel detector layers can be interpreted as \textit{RGB channels} of the same 2D picture (i.e. an additional dimension). The inputs are fixed-size windows of pixel (the jet core regions). The features inside the filters are the track patterns on the four layers thus the filters dimension must be large enough to include the track hits on the four layers. The network is realized with convolutional layers only: a 2D-picture output allows to be completely independent on the number of tracks in the layer but only to the mean occupancy. In addition, the network can be rescaled for different window size or different tracker geometry without changing the architecture but few hyper-parameters only. Another relevant feature of the convolutional approach is that all the seeds are predicted at the same time, an not removing the correspondent hits in a sequential way. This approach take has been previously used in Ref. \cite{yolo:v3} to identify a variable number of targets in videos with a real time detection.

\subsection{DeepCore Neural Network}

\begin{figure}[!htb]
  \centering
  \subfloat[]{\includegraphics[width=0.2\linewidth]{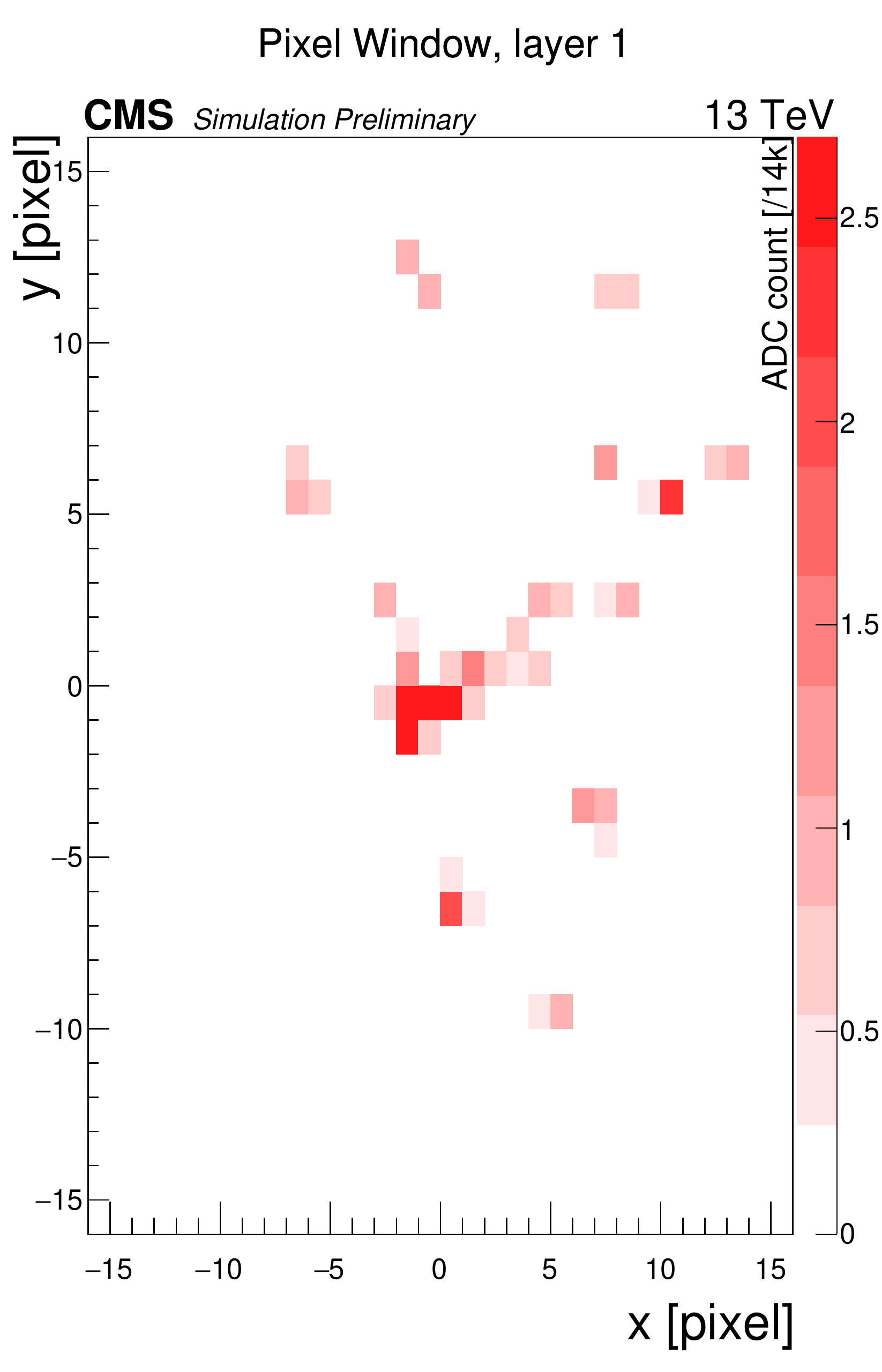}}
 \quad
  \subfloat[]{\includegraphics[width=0.2\linewidth]{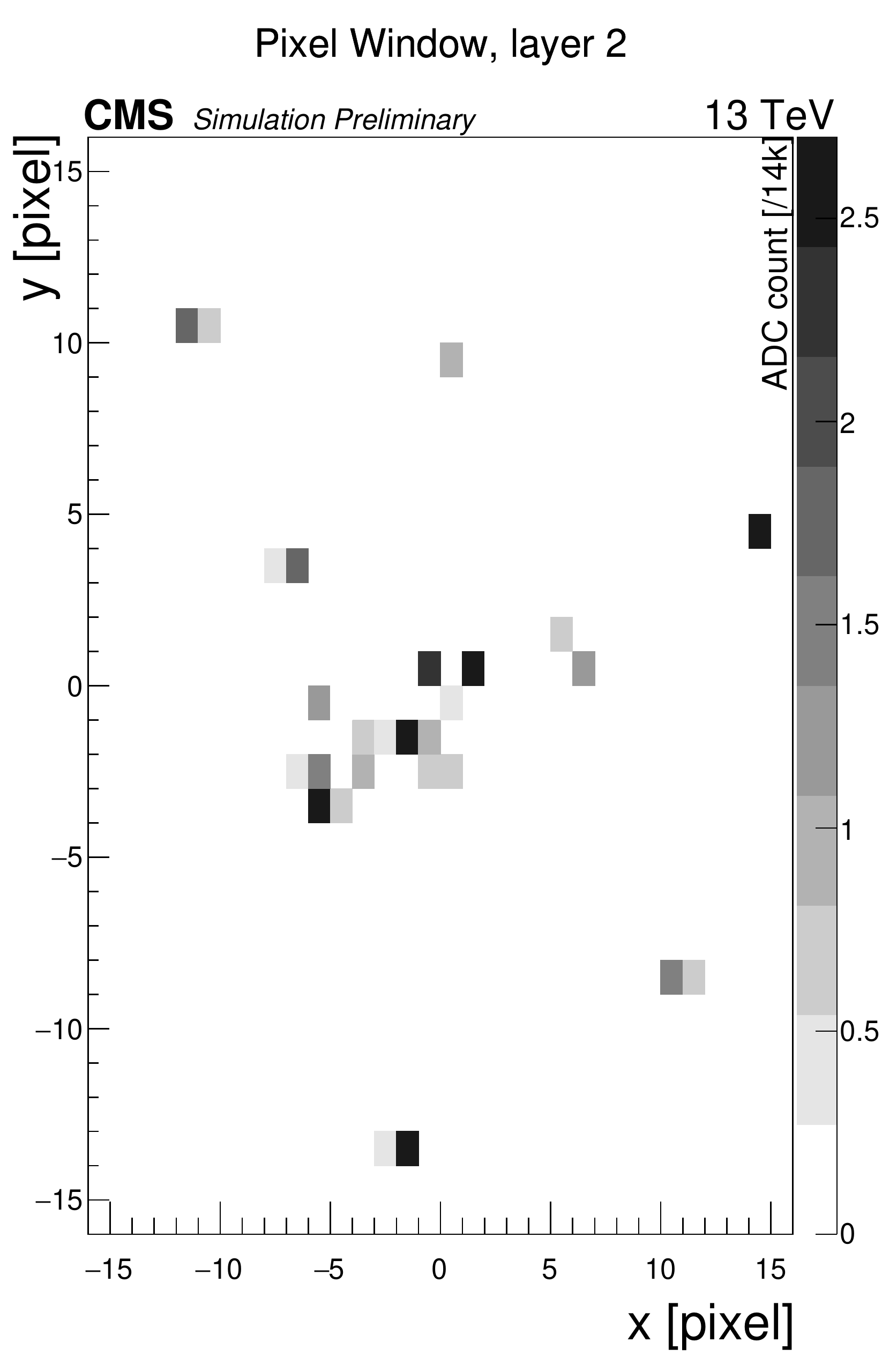}}
   \quad
    \subfloat[]{\includegraphics[width=0.2\linewidth]{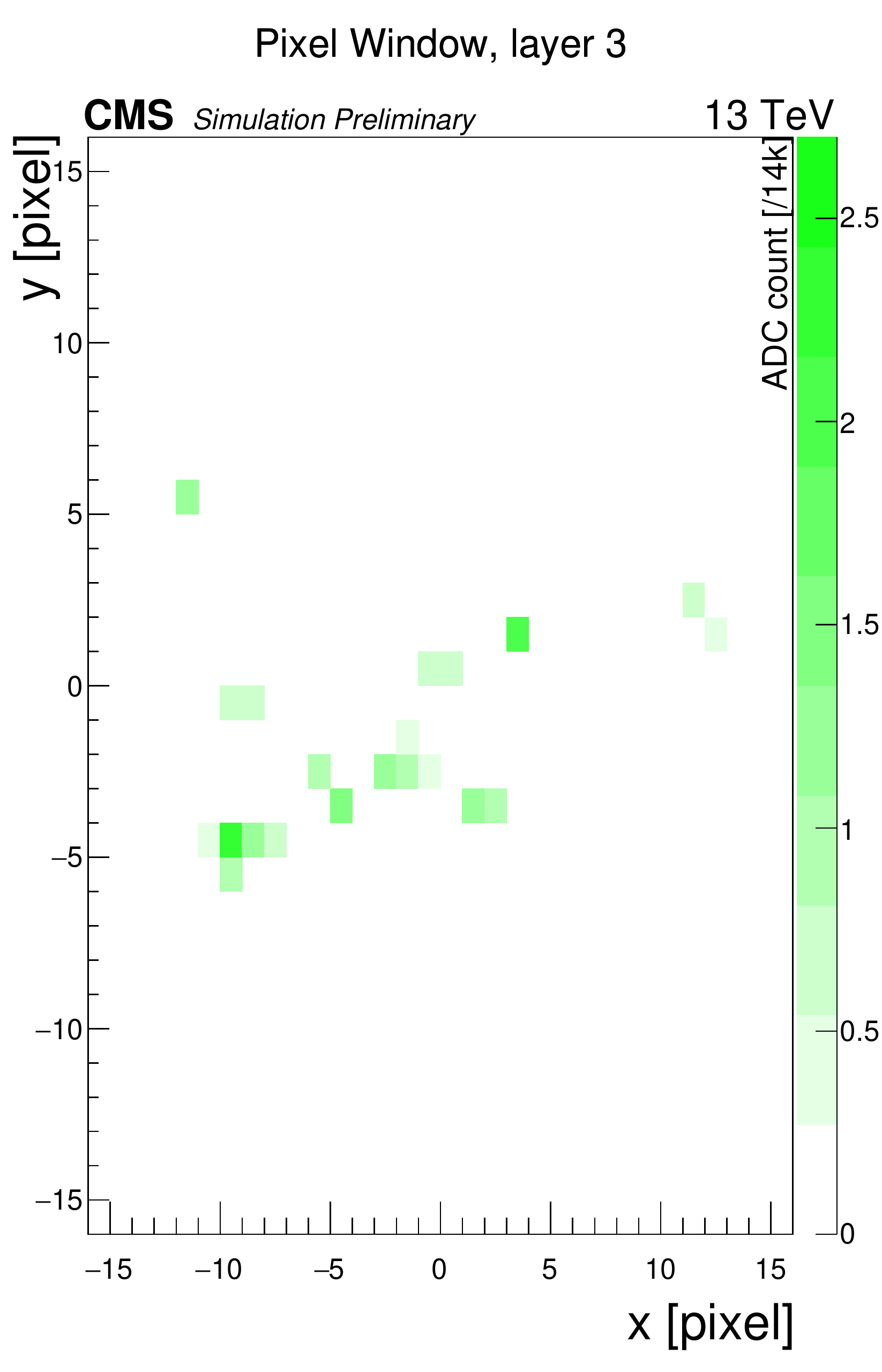}}
     \quad
  \subfloat[]{\includegraphics[width=0.2\linewidth]{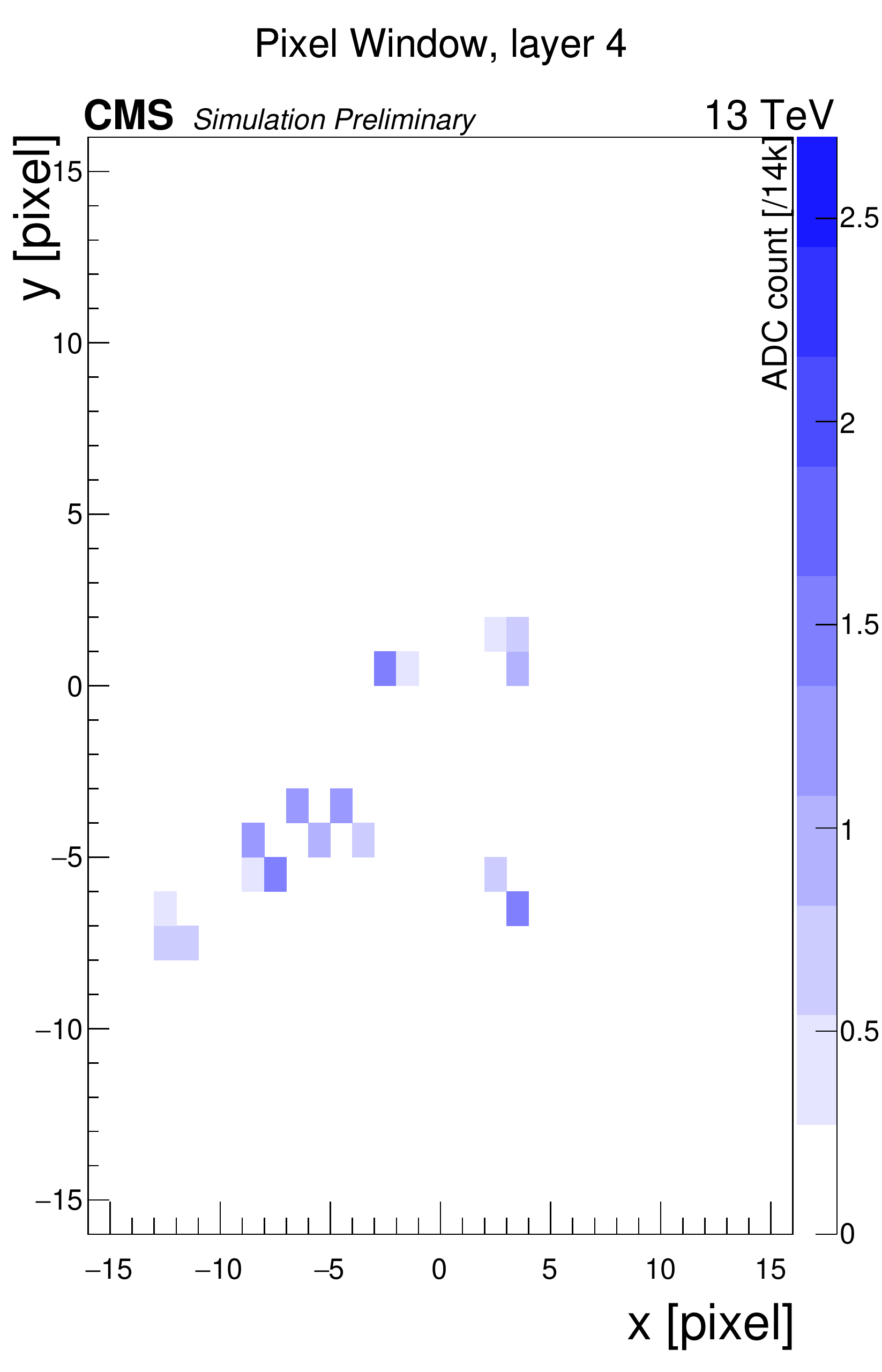}}
  \caption{Example of of the pixel maps used as input for the DeepCore neural network: the maps shows a windows on the four pixel detecor layer of CMS, aligned to the jet direction. The ADC counts are divided by 14000.}
  \label{fig:input}
\end{figure}

\begin{paragraph}{Training Input.} The input of the network are four pixel \textit{maps} centred on the merged clusters. The procedure to build them is: for each jet with $p_T>1$ TeV the interception between the jet axis from the calorimeter information and the first layer of pixel detector is found, then it is opened a cone of $\Delta R=0.1$ and are found all the merged clusters inside the cone on the layer 1. A cluster is flagged as \textit{merged} if its charge and shape are compatible with multiple particles\footnote{This assumption is used for the training input only and does not bias the CNN with respect to an MC-truth merged cluster because of the large overlap between windows.}. If the crossed pixel detector module is inactive the list of the merged cluster on the next layer, layer 2, is used. Then, for each merged cluster a $30\times 30$ pixels window is opened in each of the four layers, using as a center the interception between the layer and the direction defined by the primary vertex (PV) and the merged cluster. Also the jet axis is added as additional direction to open the four windows. For each of the direction, for each window, the $x,y$ and charge information of the hits inside the windows is stored. The charge information is normalized to a fixed value\footnote{14000, the mean value of the charge deposition in a pixel.} to obtain an ADC count number of order 1, easier to handle for a NN. Each training input is made of the four windows, called pixel maps, thus for each jet multiple overlapping inputs are produced. In addition also the jet $\eta$ and jet $p_T$ are given as input, because the shape of the cluster depends on the energy and the crossing angles of the particles. In Figure~\ref{fig:input} an example of the four pixel maps input is shown.
\end{paragraph}

\begin{paragraph}{Training Target.} For each input the target of the network is made of three copies of a \textit{Track Crossing Points (TCP) Map} and a \textit{Track-Parameters Map}. Each copy of the two Maps is pair of $30\times 30$ matrices. For each pixel of layer 2 input map, if a track crosses that pixel, 
1 will be stored in the correspondent pixel of the first TCP Map, 0 will be stored in the pixel otherwise. For each 1-pixel of the TCP Map the track parameters of the track are stored in the correspondent pixel of the Track-Parameters Map. The track parameters are stored in local coordinate: $\Delta x$ and $\Delta y$ with respect to the center of the pixel, $\Delta \eta$ and $\Delta \phi$ with respect to the merged cluster-PV direction and the $p_T$ of the track. The track parameters are also stored for the pixels in a radius of 2 pixels with respect to each TCP, with the local pixel reference (these pixels has been called Near to track Crossing Points, NCP). The rest of the pixels of the Track-Parameters Map are filled with 0. The second and third copies (called Overlap 2 and 3 Maps) are filled to take into account of multiple tracks which cross the same pixel: if another track crosses a TCP another 1 will be is stored in the TCP Map-Overlap 2 with the relative filling for the Track-Parameters Map-Overlap 2. The same for the overlap 3 Maps in case of three tracks in the same pixel. In Figure~\ref{fig:target_architecture} this complex target is shown graphically. 
\end{paragraph}


\begin{paragraph}{Architecture.}

The architecture of the network is completely convolutional. It is schematically shown in Figure~\ref{fig:target_architecture}. The inputs feed five 2D convolutional layers with reducing filter size and number, then the  network is split in two trunks: four 2D convolutional layers to produce the Track-Parameters Maps and four 2D convolutional layers for the TCP Maps. The activation functions are ReLU for all the layer but the last TCP Maps layer, where Sigmoid is used\footnote{Sigmoid is recommended for limited range output and binary losses. See training details later on.}. The total number of parameters of the network is 77373. 
\end{paragraph}

\tikzstyle{block} = [rectangle, draw, fill=lightblue!25, text width=12em, text centered, rounded corners, minimum height=4em, node distance = 2.5cm]
\tikzstyle{line} = [draw, -latex']
\tikzstyle{inout} = [draw, ellipse, fill=green!20, node distance=2cm, text width=8em, text centered]

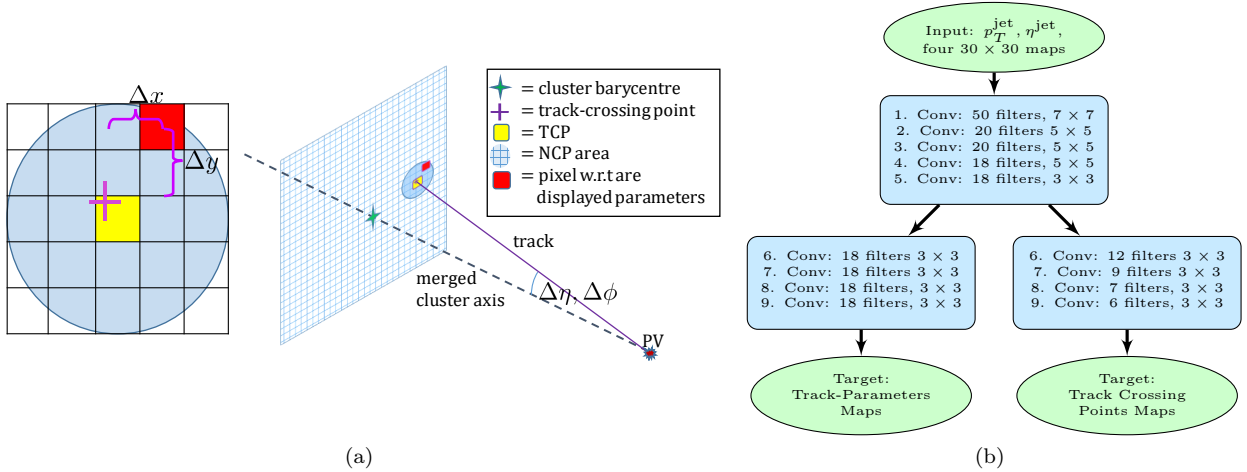
\begin{figure}[!htb]
  \centering
  \subfloat[]{\raisebox{1cm}{\includegraphics[width=0.6\linewidth]{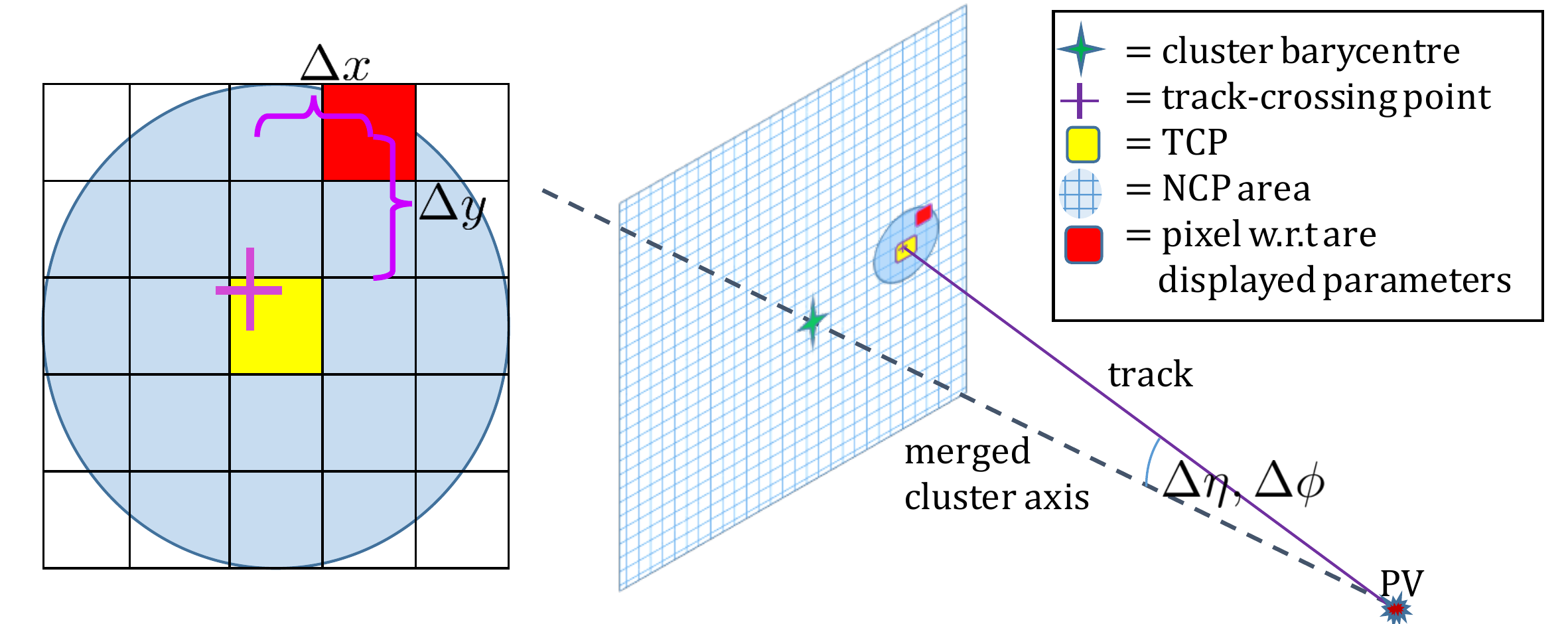}}}
  \subfloat[]{
  \tiny{
\begin{tikzpicture}[auto]
\node[inout] (input){Input: $p_T^\text{jet}, \eta^\text{jet}$, four $30\times 30$ maps };
\node [block, below of  =input, node distance = 1.5 cm] (common) {\textcolor{lightblue!25}{...}
\\1. Conv: 50 filters, $7\times 7$
2. Conv: 20 filters $5\times 5$
3. Conv: 20 filters, $5\times 5$
4. Conv: 18 filters, $5\times 5$
5. Conv: 18 filters, $3\times 3$\\
\textcolor{lightblue!25}{...}
};
\node [block, below left of=common](par){\textcolor{lightblue!25}{...}
\\6. Conv: 18 filters $3\times 3$
7. Conv: 18 filters $3\times 3$
8. Conv: 18 filters, $3\times 3$
9. Conv: 18 filters, $3\times 3$\\
\textcolor{lightblue!25}{...}
};
\node [block, below right of=common](prob){\textcolor{lightblue!25}{...}
\\6. Conv: 12 filters $3\times 3$
7. Conv: 9 filters $3\times 3$
8. Conv: 7 filters, $3\times 3$
9. Conv: 6 filters, $3\times 3$\\
\textcolor{lightblue!25}{...}
};
\node[inout, below of =prob, node distance = 1.5cm] (outprob){Target: \\ Track Crossing Points Maps};
\node[inout, below of =par, node distance = 1.5cm] (outpar){Target: \\ Track-Parameters\\ Maps};
\path [line, very thick] (input) -- (common);
\path [line,  very thick] (common) -- (prob);
\path [line,  very thick] (common) -- (par);
\path [line,  very thick] (par) -- (outpar);
\path [line,  very thick] (prob) -- (outprob);
\end{tikzpicture}}
} 
  \caption{On the left (a) an example of the Target for a single track: on the left the TCP (in yellow), the track parameters are stored for all the pixels inside the shaded blue area, the red pixel is the one with respect of which are evaluated the parameters. The Overlap Maps are not shown. On the right (b) the architecture of DeepCore.}
  \label{fig:target_architecture}
  \vspace{-0.4 cm}
\end{figure}
\begin{paragraph}{Prediction.} The Prediction of the network has the same structure of the Target i.e. three $30\times 30$ TCP Maps and three Track-Parameters Maps. The TCP Maps will contains values between 0 and 1 for each pixel thus can be interpreted as a probability that a track cross that pixel. The Track-Parameters Maps contains instead the five parameters for the TCP and NCP pixels in local coordinate. 
\end{paragraph}

\begin{paragraph}{Training details.} The NN has been trained with a large sample of inputs, for which also the relative target information is given. During the training the network must predict the target given the input only, then it must compare the prediction with the true target. The comparison proceeds with a given metric  i.e. the \textit{loss function}, which defines the grade of accuracy of the prediction. Two losses, one for each target, has been used to train DeepCore. A weighted \textit{Binary Cross Entropy} has been used for the TCP Maps i.e $\mathcal L_{TCP}=\frac{1}{N}\sum_{i=1}^N \Bigl[y_{i}^\text{true}\ln(y_i^\text{pred})+(1-y_i^\text{true})\ln(1-y_i^\text{pred})\Bigl]$, where the TCP-pixels have weight 10, the NCPs 1 and the other pixels 0.01. The weighting is needed to avoid a vanishing TCP prediction because of the sparse target.  A clipped mean square error has been used for all the parameters $\mathcal L_\text{par}=\frac{\sum_{p\in TCP,NCP} \min[(p_\text{pred}-p_\text{true})^2,25]}{N_{TCP+NCP}}$, where the sum runs only on the TCP and NCP pixels. The clipping is needed to avoid large tails in the prediction which enlarge artificially the loss. The training sample is composed by 22 millions of input (about 2 millions of jets) plus two million used for validation and it is composed of multijet events with the trasfer $\hat p_T$ between 1.8 and 2.4 TeV. The jets are required to have $p_T^\text{jet}>1$ TeV and $|\eta^\text{jet}|<1.4$, while only the tracks with $p_T>1$ GeV has been used to build the targets. The batch size (the number of input analysed for each prediction) is 32, which is the largest possible given the available computation power. The chosen optimizer is Adam \cite{adam}, the learning rate has been changed during the 246 epochs of training, gradually from $2\cdot 10^{-4}$ to $10^{-7}$, and in each epoch all the training sample is explored. 
\end{paragraph}

\subsection{Integration of DeepCore in CMS reconstruction}

The training of DeepCore has been performed outside of the CMS reconstruction software (CMSSW) on GPU and then the final weights have been permanently stored and given to CMSSW. DeepCore has been developed with Keras library \cite{keras} both for the training and the prediction inside CMSSW. DeepCore has been integrated into the jetCore iteration of CMS reconstruction: standard jetCore seeding has been disabled and the following algorithm is the replacing. 

The cluster list in a cone of $\Delta R=0.1$ with the respect of the jet axis is identified for each calorimeter jet with $p_T>300$ GeV. Each cluster defines a new direction on which a DeepCore Input is built (the four pixel maps and the $p_T^\text{jet},|\eta^\text{jet}|$). The input is defined for all the cluster and not for the merged cluster to recover as much efficiency as possible at seeding level, the standard duplicate remover will take into account to remove overlapped tracks in the following steps of reconstruction. The input is given to DeepCore NN which returns the prediction given the weights of the training. The list of actual seeds is made from DeepCore prediction with the sets of five track parameters of the \textit{most probable} pixels. \textit{Most probable} is defined as TCP output greater than 0.85, 0.75, 065 for the three Overlaps or greater than 0.5, 0.4, 0.3 in case the layer 2 is missing for the given input (because of inactive module). The threshold is lowered in the latter case because the target of TCP is built on layer 2, thus it is crucial in the prediction. In addition to the standard duplicate remover, the list of seeds is cleaned from duplicates: if two seeds have $\Delta x,\Delta y < 50$ $\mu$m, $\Delta \eta, \Delta \phi<0.002$ the one from the lower value of TCP is removed from the list. The uncertainty on the parameters is fixed for all the seeds: $\sigma_{p_T}=0.15$ GeV, $\sigma_\eta=\sigma_\phi=0.01$, $\sigma_{xy}=\sigma_z=44$ $\mu$m, without off-diagonal terms, based on the performance of the prediction of DeepCore (see next section). 

\section{Preliminary performance of DeepCore}

\begin{figure}[!htb]
  \centering
  \subfloat[]{\includegraphics[width=0.2\linewidth]{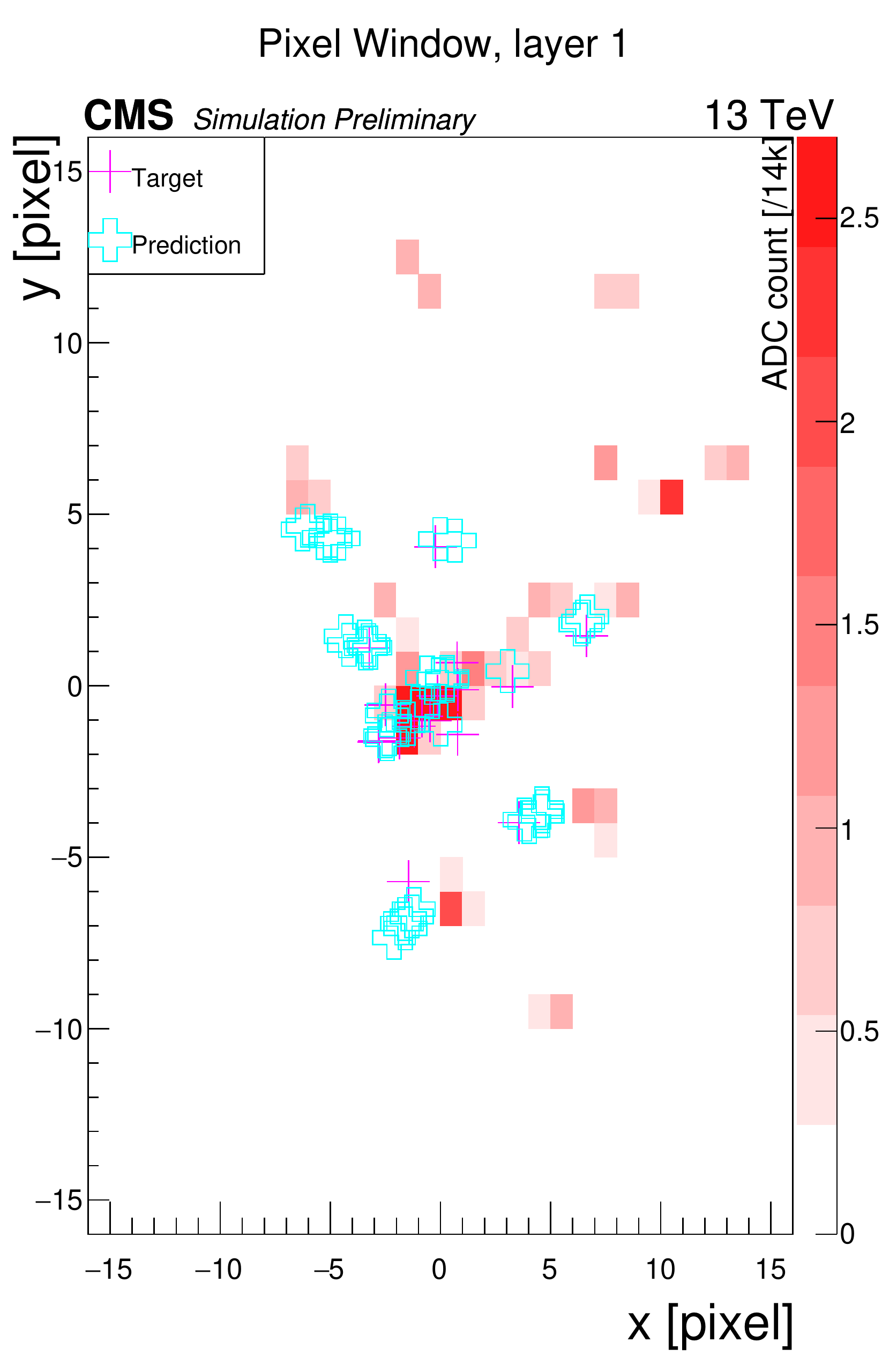}}
  \subfloat[]{\includegraphics[width=0.2\linewidth]{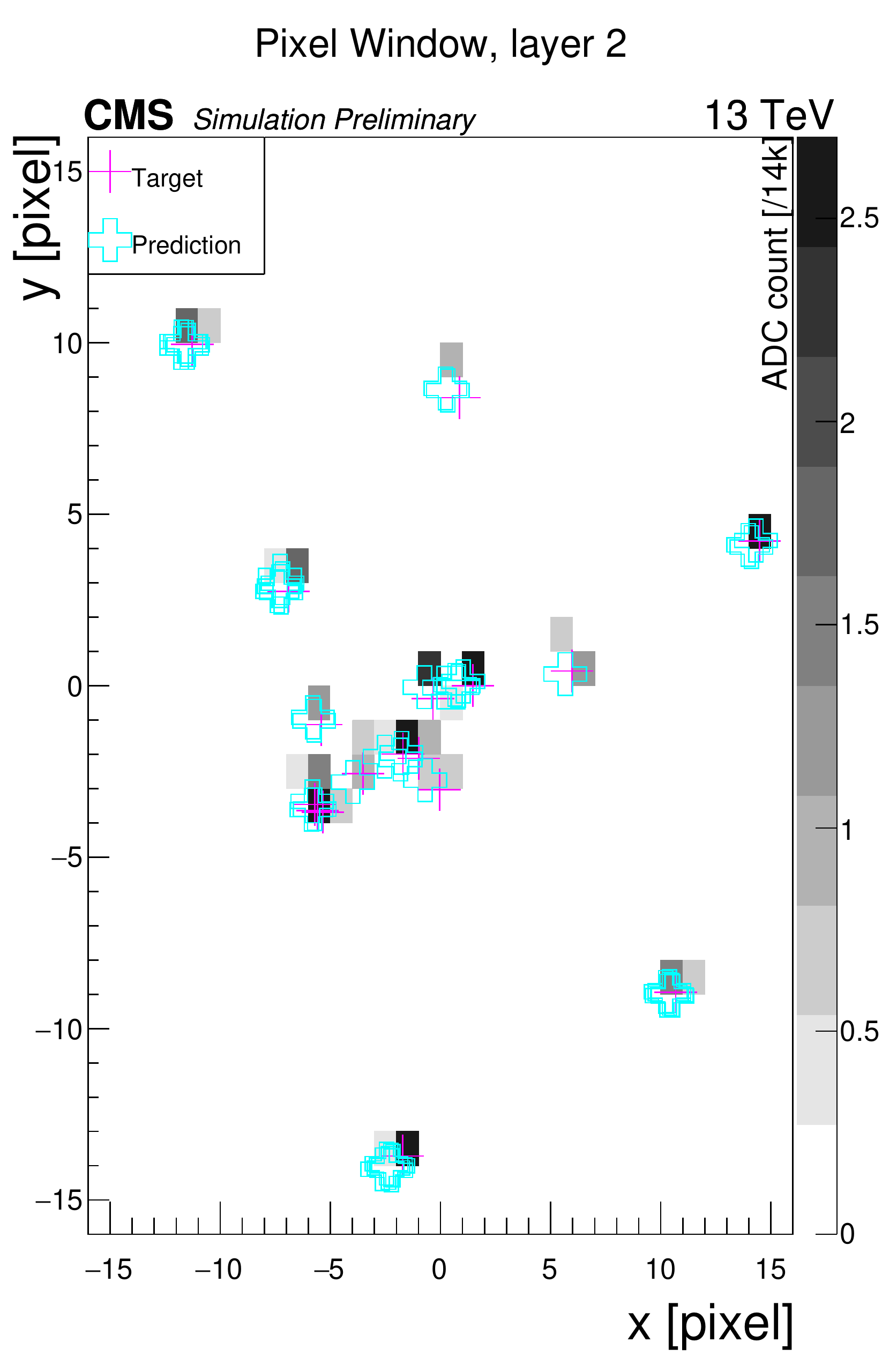}}
    \subfloat[]{\includegraphics[width=0.2\linewidth]{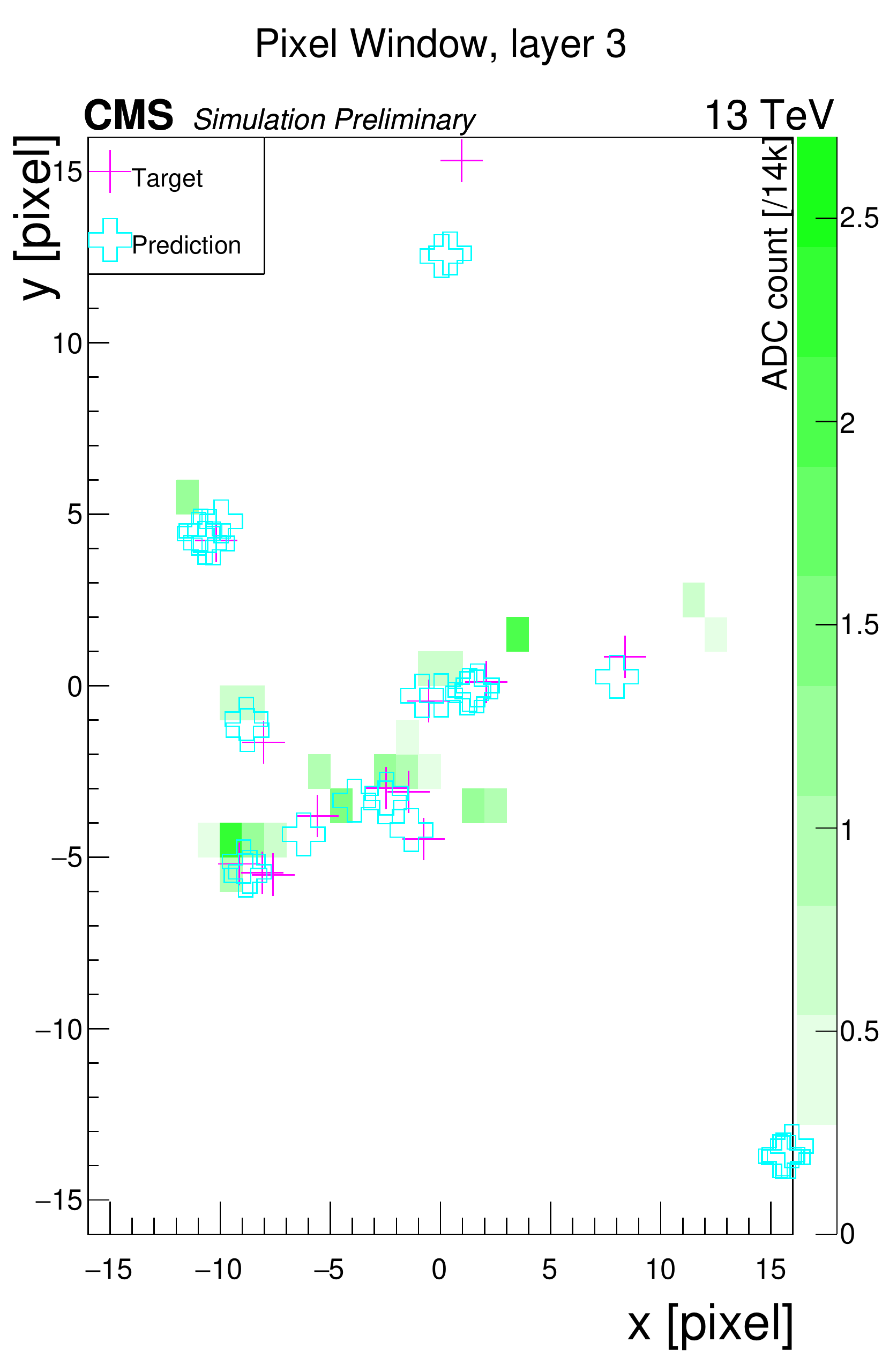}}
  \subfloat[]{\includegraphics[width=0.2\linewidth]{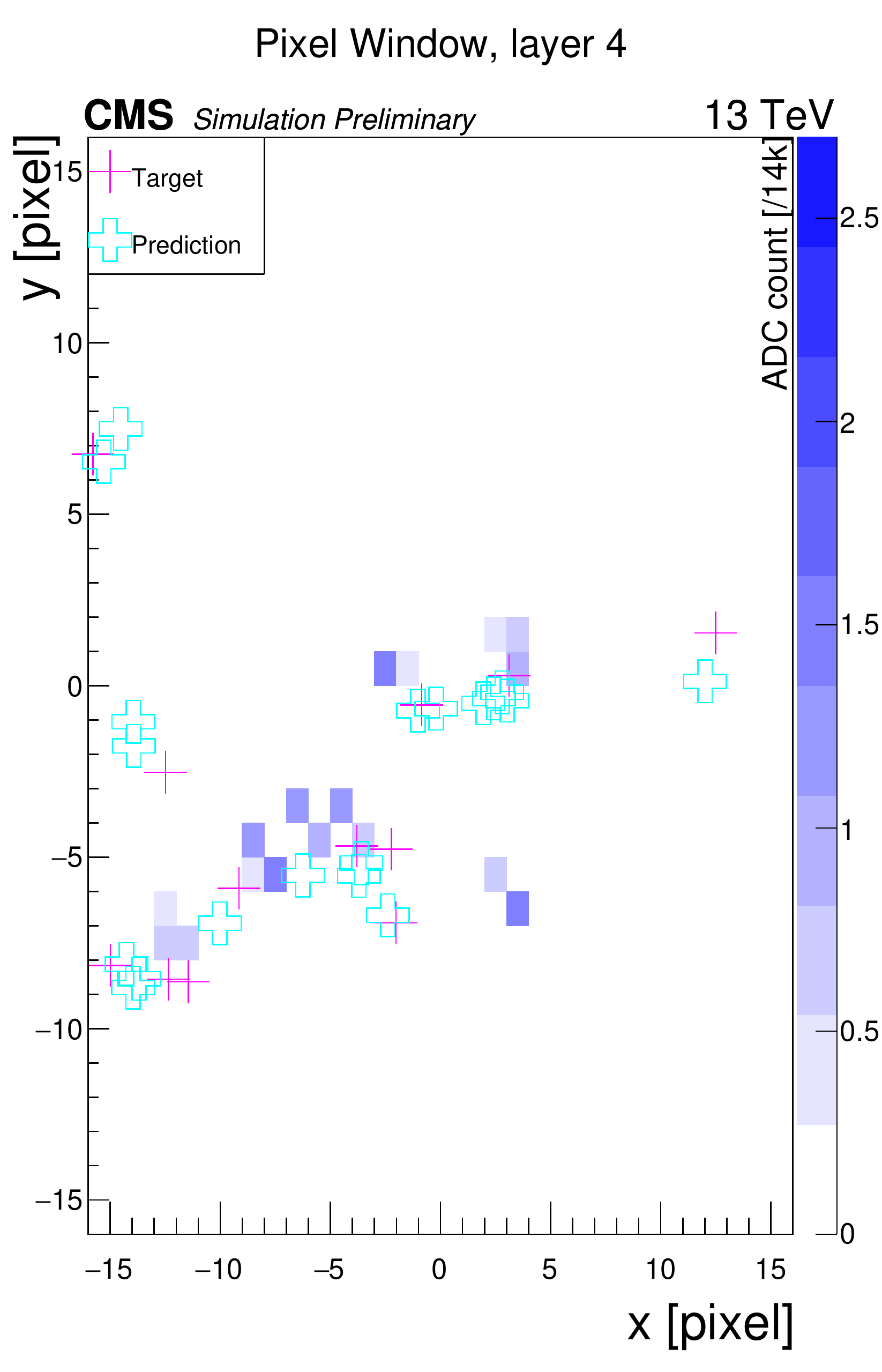}}
  \subfloat[]{\includegraphics[width=0.21\linewidth]{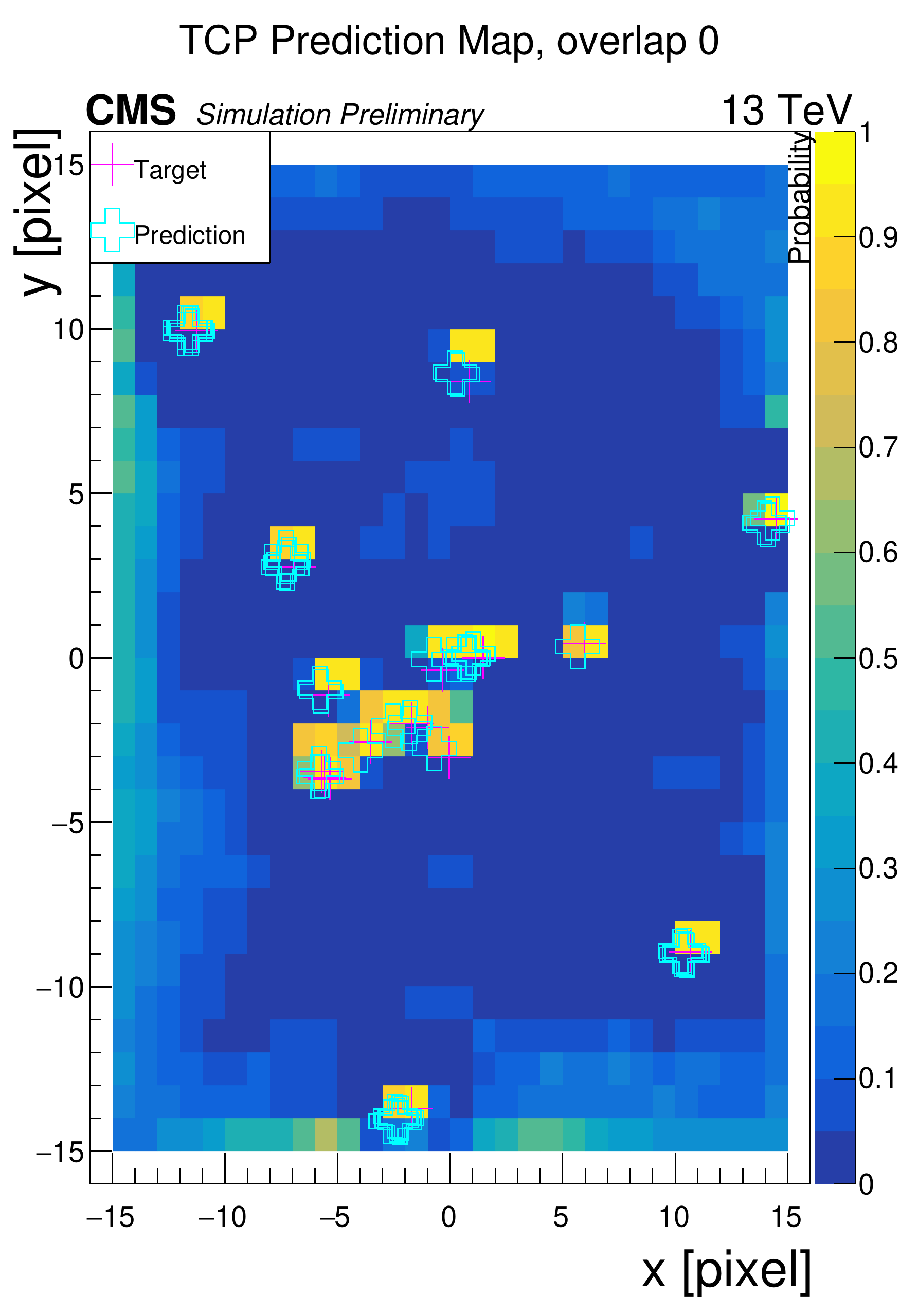}}
  \caption{Example of the pixel maps used as input. On the top are also shown the crosses of the crossing point of the target (simulated) tracks and the correspondent prediction of DeepCore for the most probable hits. The prediction is produced on layer 2 and propagated linearly on the other layers. The most right figure is the map of the predicted crossing point on the window of layer 2, expressed as probability, with the crosses of the predictions and the targets. The linear propagation is used in the event display only, in seed production the predicted $p_T$ is used.}
  \label{fig:event_display}
\end{figure}

\begin{figure}[!htb]
  \centering
  \subfloat[]{\includegraphics[width=0.45\linewidth]{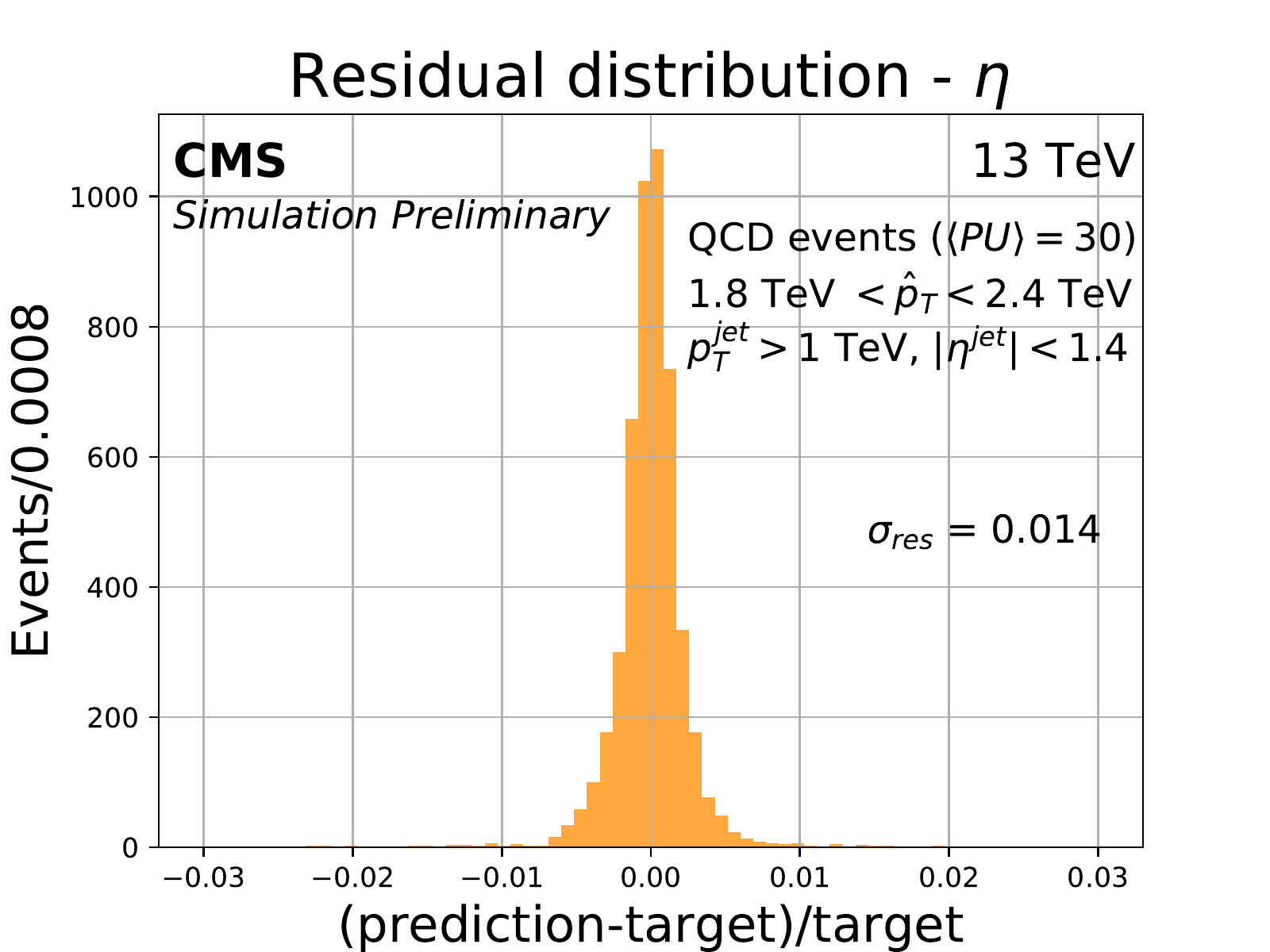}}
  \qquad
  \subfloat[]{\includegraphics[width=0.45\linewidth]{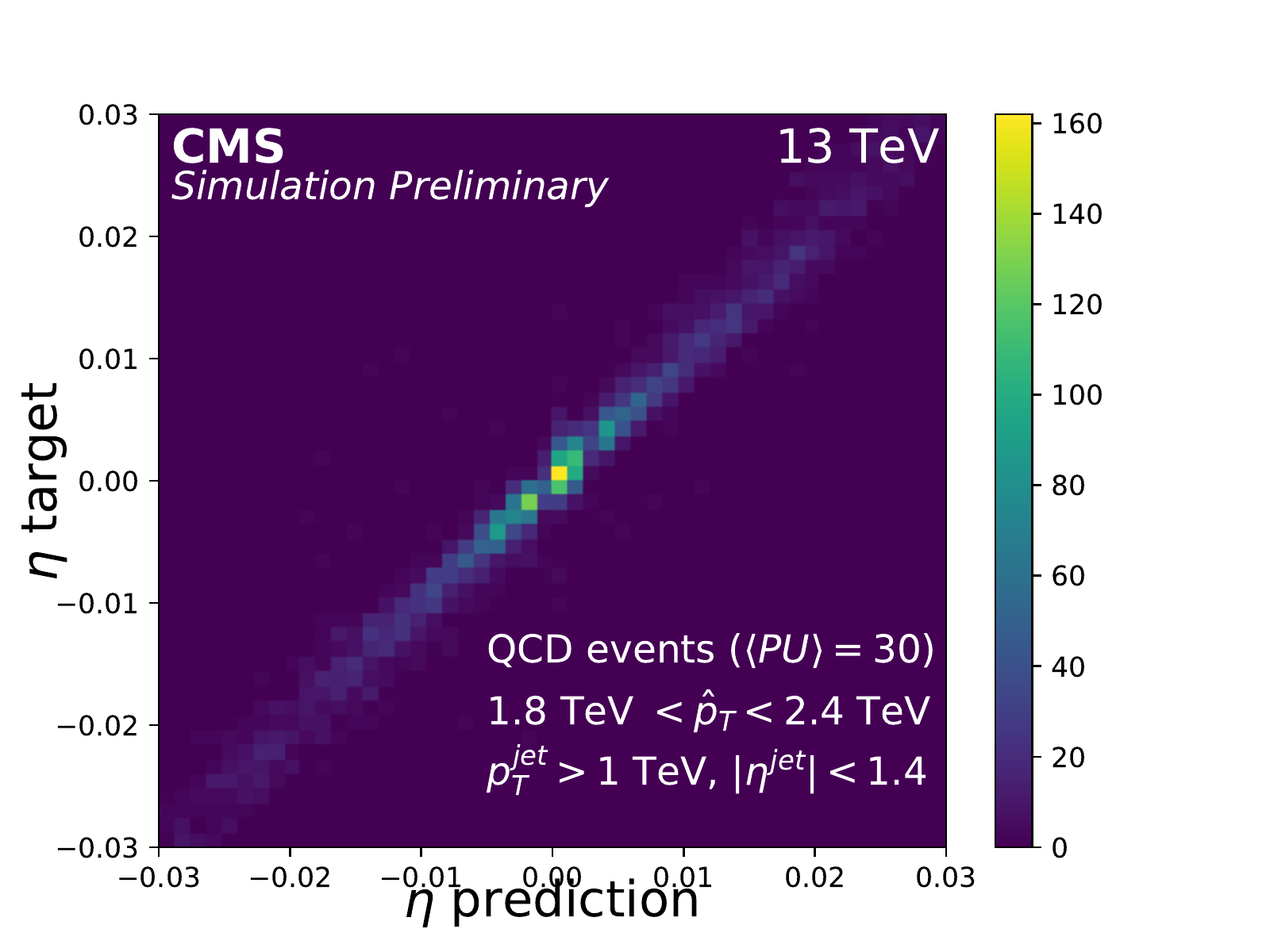}}
  \caption{ On the left (a) the residual between the seed $\eta$ parameter predicted by DeepCore and the target (simulated) track $\eta$ parameter. On the right (b) the correlation between prediction of DeepCore and target parameters shown with seed $\eta$ parameter predicted against the simulated track $\eta$ parameter.}
    \label{fig:residuals}
\end{figure}

The behaviour of DeepCore can be checked during the training with an "event display", developed for optimization studies externally from CMSSW. The same event of Figure~\ref{fig:input} is shown in Figure~\ref{fig:event_display}, together with TCP Map, the target and the track-parameter prediction of the \textit{most probable} hits only, at the end of the training. The event display has only a qualitative interpretation, but it reveals an almost full efficiency and an accuracy of 1-2 pixels also with the used linear propagation, with an affordable level of duplication (the duplicate remover has not been run here).  The Figure~\ref{fig:residuals} shows an example of the quantitative validation of the training performance, in term of residual of $\eta$ parameter between the prediction and the target. The null average bias, the 1.4\% spread and the strong correlation with the target show that DeepCore is able to predict the parameters given the pixel input.  

DeepCore has been validated integrated in the CMS renconstruction on 20k multijet events  with the trasfer $\hat p_T$ between 1.8 and 2.4 TeV. The jets are required to have $p_T^\text{jet}>1$ TeV and $|\eta^\text{jet}|<1.4$ and on the simulated tracks has been applied the typical CMS selection $|\eta|<2.5$, $r_\text{prod}<3$ cm, $|z_\text{prod}|<30$ cm, $p_T>0.9$ GeV.

\begin{figure}[!htb]
  \centering
  \subfloat[]{\includegraphics[width=0.45\linewidth]{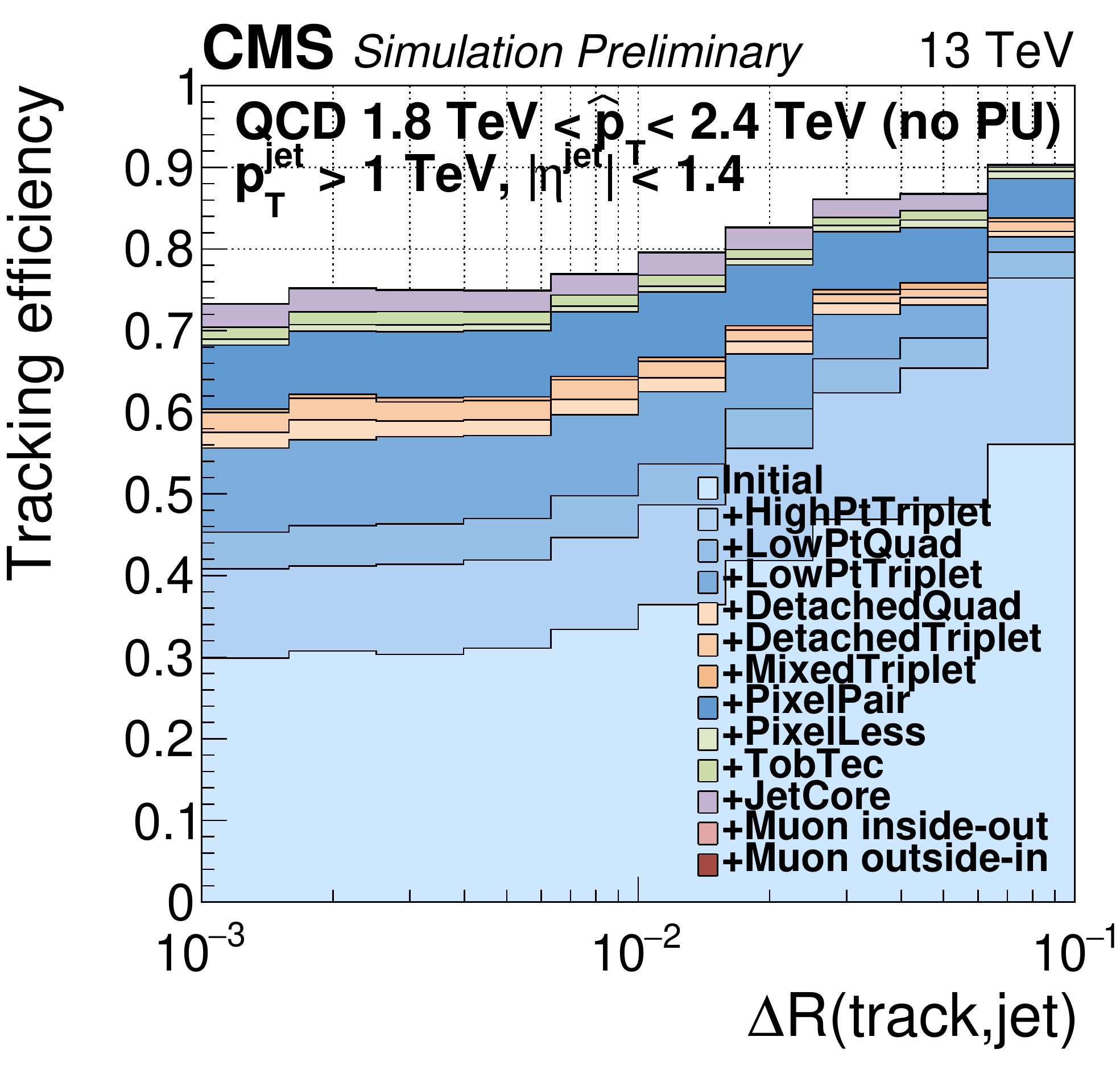}}
  \quad
  \subfloat[]{\includegraphics[width=0.45\linewidth]{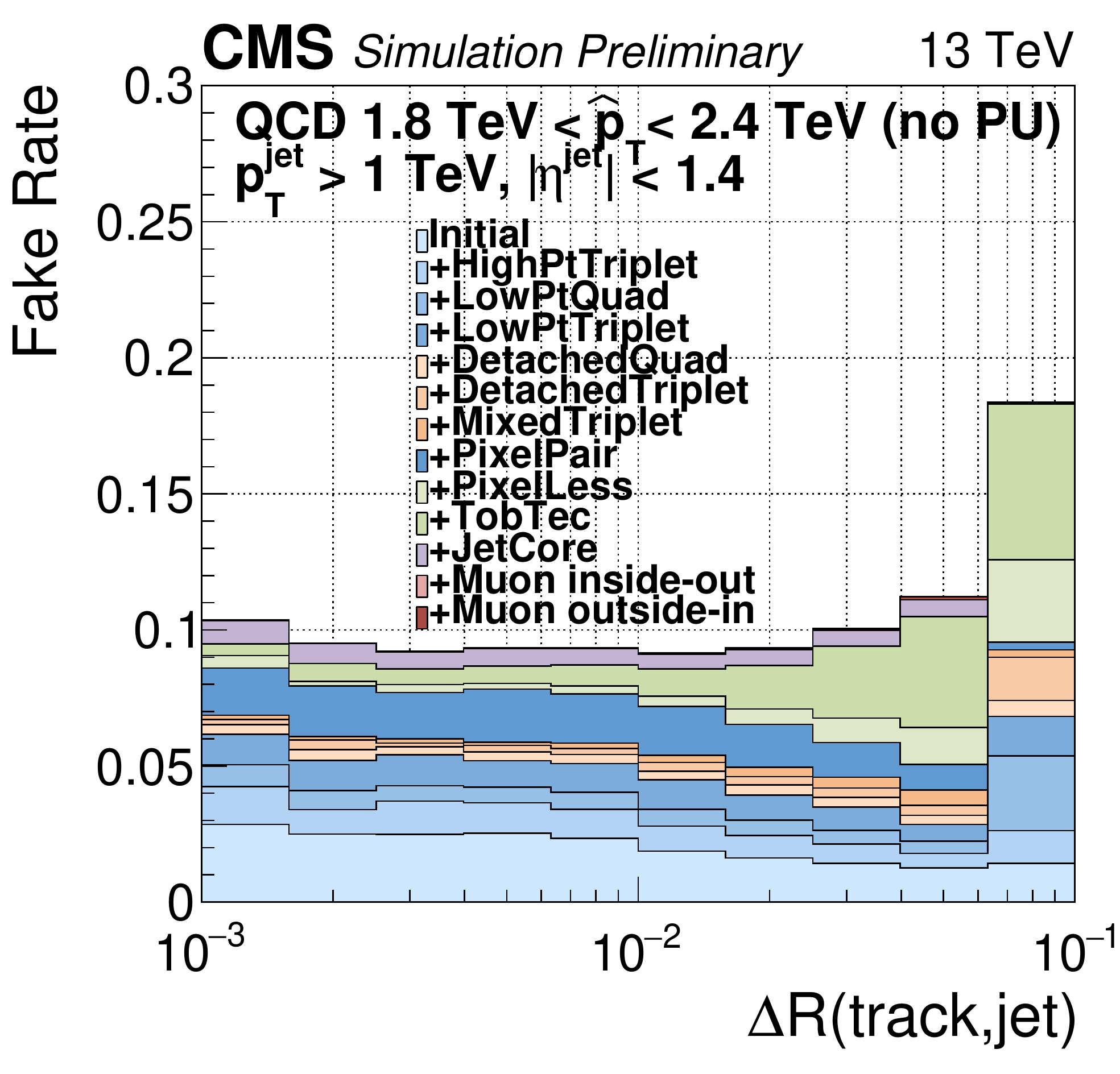}}
  \caption{Tracking efficiency (left figure) and fake rate (right figure) in the jet core region ($\Delta R<0.1$, between the reconstructed jet axis and the simulated track direction). The contribution of the different iterations of the CKF are shown as stacked histograms. The DeepCore algorithm is used in the iteration dedicated to the cores of the jets [jetCore (purple)]. In the efficiency the shared reconstructed tracks (duplicated) between various iterations are not removed.
}
    \label{fig:stacked}
\end{figure}

The result for the final tracking performance are shown in Figure~\ref{fig:stacked} in a stacked plot with highlighted the contribution of the various iterations of the CTF in the jet core region.
The tracking efficiency is defined as $\varepsilon=N_\text{assoc}/N_\text{sim}$, where $N_\text{sim}$ is the number of simulated tracks and $N_\text{assoc}$ is the number of simulated tracks associated to a reconstructed one. The fake rate is defined as $R_F=N_\text{not assoc}/N_\text{reco}$, where $N_\text{reco}$ is the number of reconstructed tracks and $N_\text{not assoc}$ is the number reconstructed tracks not associated to a simulated one. A reconstructed track is flagged as “associated” if the $\chi^2$ between its parameters and the simulated is lower than 25. This definition replaces the usual CMS one (based on the fraction of true hits used) for these validation studies, because DeepCore seeding is without pixel hits and with the usual association it will be negatively biassed. 

\begin{figure}[!htb]
  \centering
  \qquad
  \subfloat[]{\includegraphics[width=0.45\linewidth]{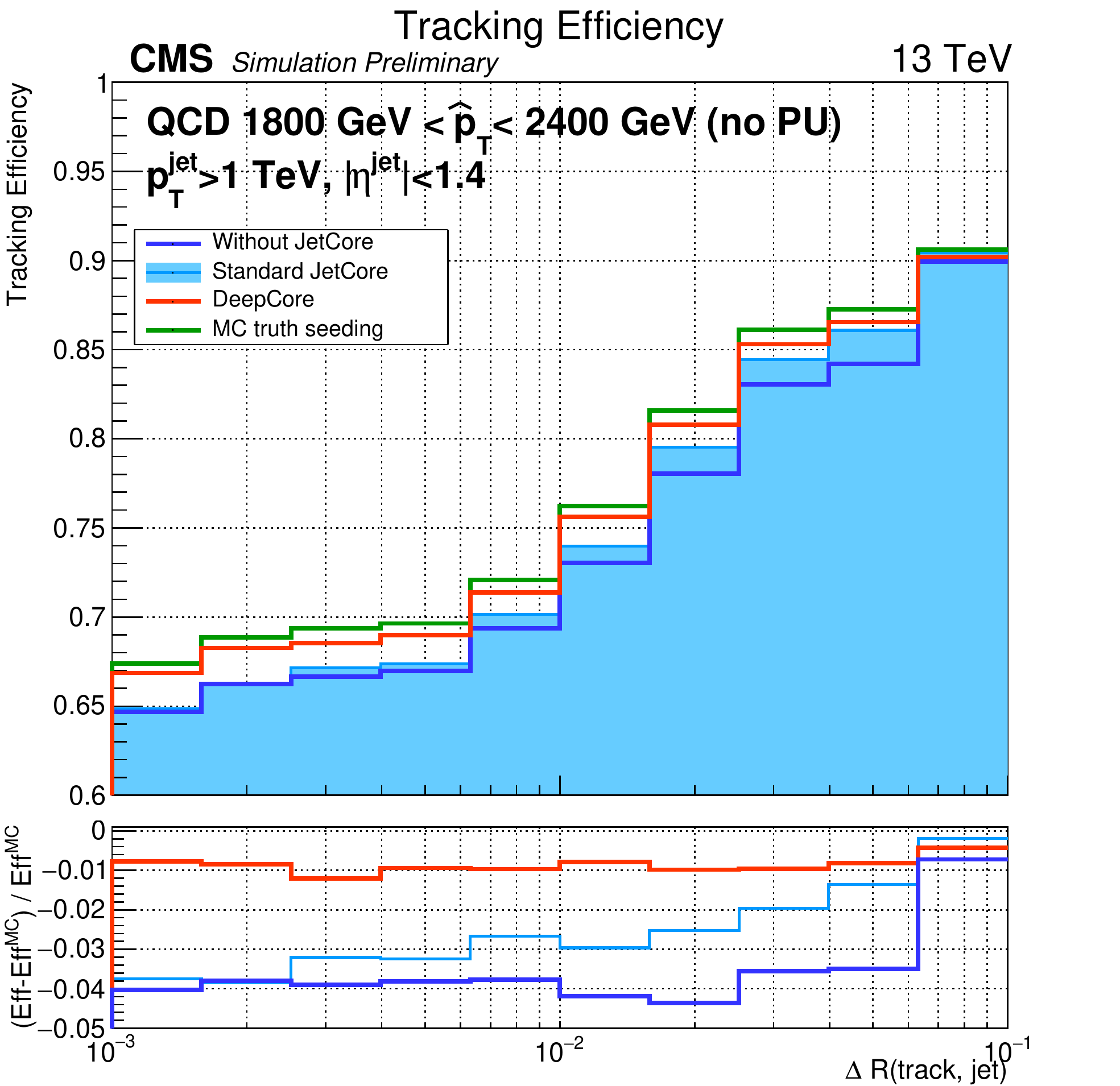}}
  \qquad
  \subfloat[]{\includegraphics[width=0.45\linewidth]{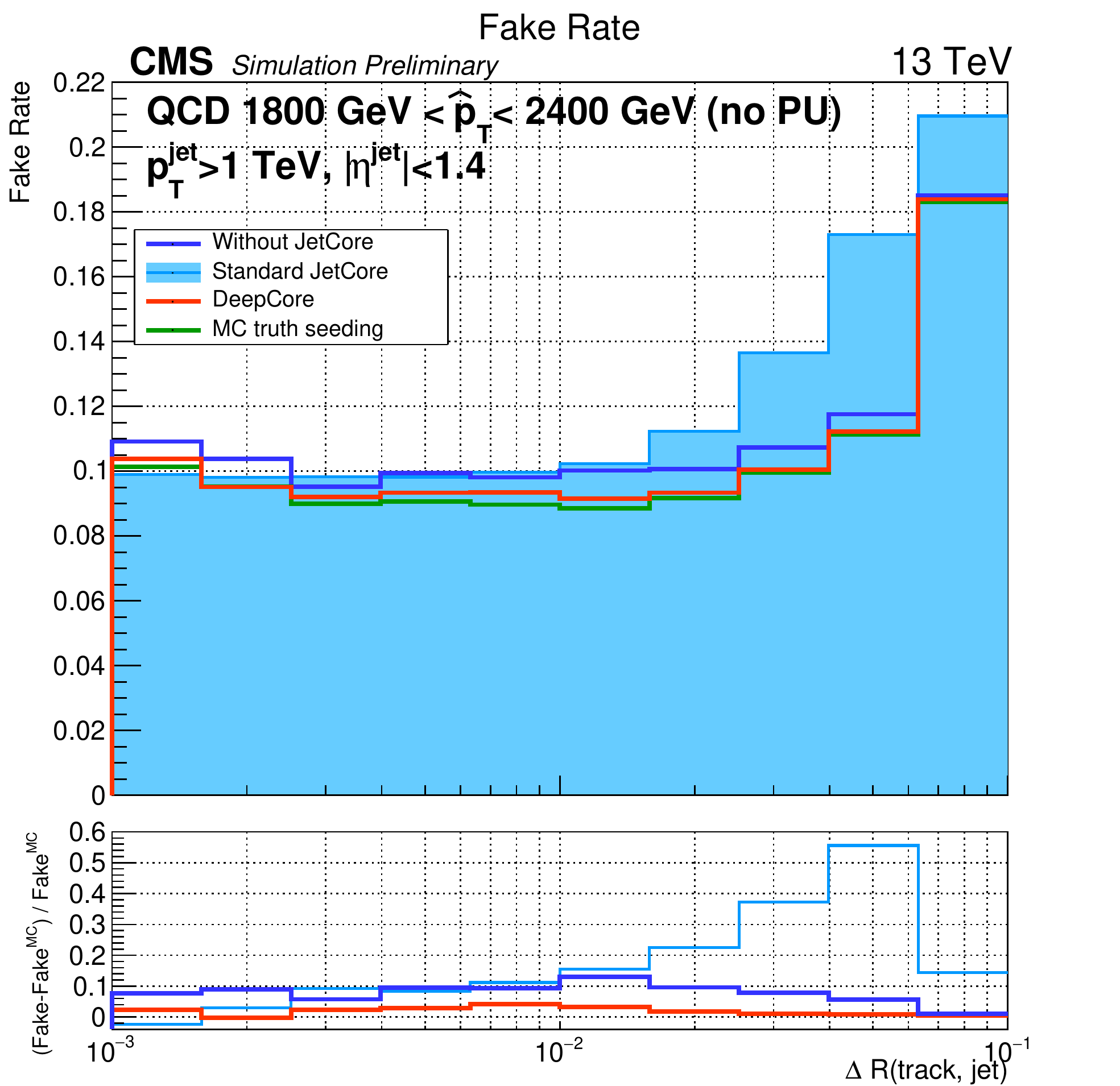}}
  \caption{Tracking efficiency (left figure) and fake rate (right figure) in the jet core region ($\Delta R<0.1$, between the reconstructed jet axis and the simulated track direction). The light blue filled histogram is obtained with the standard CMS tracking algorithm. The dark blue histogram is obtained removing the CKF iteration dedicated to the jet cores. The red histogram is obtained using the DeepCore in the seeding for the iteration dedicated to the jet cores. The green histogram is obtained producing the seed for the jetCore iteration using the MC truth seeding. In the lower pads are shown the differences between various tracking efficiencies (fake rates) and the MC truth seeding one, divided by the MC truth seeding efficiency (fake rate).}
    \label{fig:compared_perf}
\end{figure}

The improvement given by DeepCore to CMS reconstruction is better shown in Figure~\ref{fig:compared_perf}, where are compared the performance with the standard jetCore algorithm and the one with DeepCore. 
Also the tracking performance obtained producing the seed for the jetCore iteration using the simulated track information is shown (\textit{MC truth seeding}), for which the seeding efficiency is 100\% and the fake rate 0\% by definition. DeepCore is able to reproduce the perfect seeding efficiency with degradation below 1\%, flat in $\Delta R$. On the other hand, all the fake tracks produced by the standard jetCore are avoided, reducing the seeding fake rate below 5\%. 
In particular the good purity of DeepCore seeds lower the fakes below the rate without the jetCore iteration because DeepCore is able to correctly reconstruct tracks reconstructed as fakes by different iteration in the low $\Delta R$ region. 

Also the timing performance has been validated: the DeepCore time consumption is 15\% of the average time of standard jetCore iteration.

\section{Conclusions}

The CNNs have been shown to be a valid approach to perform seeding for track reconstruction in a dense environment. The DeepCore algorithm, developed and validated with the CMS tracker in the central region, shows better performance than the standard seeding algorithm in such dense environment: it almost cancels the seeding inefficiencies, reduces the fake rate up to 60\% and the seeding time by 85\%.  
For the track reconstruction to be used in Run3 of LHC, CMS plans to extend this approach and make use of the endcap region as well. The more complex geometry of the endcaps require some adjustment of the network input and target. Furthermore, an optimization of the training (in terms batch size, learning rate and architecture) and of the target definition (in order to reduce the strong dependence on the layer 2) it is planned. 
In addition specific studies are required to evaluate the impact of a pixel-less seeding in the inward CKF extrapolation. 
Finally, the good performance of the DeepCore algorithm suggests to study the impact of applying such an approach to the pattern recognition as well.



\bibliographystyle{spphys}      
\bibliography{bibliografia.bib}   

\end{document}